\documentclass[12pt,preprint]{aastex}

\begin{document}

\shorttitle{Beaming in Black Hole Binaries}
\shortauthors{Narayan \& McClintock}

\title{Inclination Effects and Beaming in Black Hole X-ray Binaries}

\author{Ramesh Narayan and Jeffrey E. McClintock}
\affil{Harvard-Smithsonian Center for Astrophysics,
Cambridge, MA 02138, USA} 
\email{rnarayan@cfa.harvard.edu,jem@cfa.harvard.edu}

\begin{abstract}
We investigate the dependence of observational properties of black
hole X-ray binaries on the inclination angle $i$ of their orbits.  We
find the following: (1) Transient black hole binaries show no trend in
their quiescent X-ray luminosities as a function of $i$, suggesting
that the radiation is not significantly beamed.  This is consistent
with emission from an accretion disk.  If the X-rays are from a jet,
then the Lorentz factor $\gamma$ of the jet is $<1.24$ at the 90\%
confidence level.  (2) The X-ray binary 4U1543--47 with $i\sim21\degr$
has a surprisingly strong fluorescent iron line in the high soft
state.  Quantifying an earlier argument by Park et al. (2004), we
conclude that if the continuum X-ray emission in this source is from a
jet, then $\gamma<1.04$.  (3) None of the known binaries has $\cos
i<0.25$ or $i>75\degr$.  This fact, plus the lack of eclipses among
the 20 black hole binaries in our sample, strongly suggests at the
99.5\% confidence level that systems with large inclination angles are
hidden from view.  The obscuration could be the result of disk
flaring, as suggested by Milgrom (1978) for neutron star X-ray
binaries.  (4) Transient black hole binaries with $i\sim70-75\degr$
have significantly more complex X-ray light curves than systems with
$i\lesssim65\degr$.  This may be the result of variable obscuration
and/or variable height above the disk of the radiating gas.
\end{abstract}


\section{Introduction}

An accretion disk is intrinsically non-spherical since the angular
momentum vector of the orbiting gas specifies a unique direction.
Depending on the direction from which one views the disk --- face-on
or edge-on or something in between --- one expects differences in what
one observes.  Apart from the obvious $\cos i$ factor which determines
the projected area of the disk, $i$ being the angle between the disk
normal and the direction to the observer, there could be additional
beaming effects if the radiating gas has relativistic motion.  The
orbital motion of the gas in the disk leads to blue- and redshifts for
observers who are close to edge-on ($i\to 90\degr$), while the
radiation from a jet is relativistically boosted for observers who are
more face-on ($i\to 0\degr$).  There may also be effects due to
obscuration.  For instance, the radiation emitted by one part of the
disk may be blocked from the observer's view by other parts of the
disk; alternatively, the secondary star may do the obscuring, in which
case one expects eclipses in the light curve.

In order to study these inclination effects, we need a sample of disk
systems for which we have independent estimates of $i$.  X-ray
binaries are particularly useful in this regard.  Because the gas
accreting onto the compact star is supplied by the secondary, the
angular momentum of the material is aligned with the orbital angular
momentum of the binary system.  Thus, the accretion disk is expected
to be coplanar with the binary orbit.  The inclination of the binary
orbit with respect to the observer's line-of-sight can be estimated
from the light curve of the secondary (e.g., Orosz \& Bailyn 1997).
This inclination directly gives the inclination angle $i$ of the disk.

We focus in this paper on black hole X-ray binaries, limiting our
study to the 20 black hole systems with dynamical mass information
(McClintock \& Remillard 2004; Orosz et al. 2004; Casares et
al. 2004).  Of these 20 sources, 17 are transient black hole X-ray
novae (BHXN) and the remaining 3 are persistent sources.  We collect
from the literature estimates of the inclination angles of these
binaries and investigate their luminosities, spectra and light curves.
Our aim is to identify any patterns or systematic variations with
respect to inclination in the data.  In \S 2, we discuss relativistic
beaming effects and look for evidence of jets in certain spectral
states of black hole binaries.  Then, in \S 3 and \S 4 we discuss
obscuration effects.  We conclude in \S 5 with a summary.

\section{Jets and Relativistic Beaming}

Consider an accreting black hole with twin relativistic jets emitted
in opposite directions.  Let us make the reasonable assumption that
the jets are perpendicular to the accretion disk so that the
inclination angle $i$ defined earlier is also the angle between the
line-of-sight and the jet direction.  If the bulk velocity of the jets
is $\beta=v/c$ and the corresponding Lorentz factor is $\gamma$, then
the Doppler factors of the approaching and receding jets are given by
\begin{equation}
\delta_a = {1\over\gamma(1-\beta\cos i)}, \qquad \delta_r =
{1\over\gamma(1+\beta\cos i)}.
\end{equation}
Let the spectrum of the source have the form $S_\nu \propto
\nu^{-\alpha}$.  Then the total flux density from both jets as seen by
an observer at angle $i$ to the jet axis is given by (see Mirabel \&
Rodr\'iguez 1999)
\begin{equation}
S_{\rm obs}(i) = S_0 \left(\delta_a^{k+\alpha} + \delta_r^{k+\alpha}
\right),
\end{equation}
where $S_0$ is the emitted flux density of each jet, and the index $k$
takes either the value 2 (for a continuous jet) or 3 (for discrete
blobs).  We assume $k=2$ in what follows, both because it is more
appropriate for our problem and because it gives more conservative
results.

The curves in Figure 1 show the variation of $S_{\rm obs}$ with $i$
for different choices of $\gamma$.  The calculations assume $\alpha=1$
(photon index 2, as appropriate for quiescent BHXN), and the curves
have been normalized so as to have the same flux density at $\cos
i=0.5$.  The effect of relativistic beaming is obvious.  Especially
for $\gamma>2$, the observed flux in the limit $\cos i \to 1$ (or
$i\to 0$) is nearly two orders of magnitude larger than the flux in
the limit $\cos i\to 0$ ($i\to \pi/2$).  Given a sample of sources
with estimates of $i$, it might be possible to test for such a strong
effect.

Below we test for the presence of jets and beaming in two different
spectral states of black hole binaries.  In \S 2.1 we describe current
ideas on the nature of the low hard state and the quiescent state of
BHXN.  In \S 2.2 we consider the luminosities of BHXN in the quiescent
state and show that there is very little evidence for jet-induced
beaming in this state, consistent with earlier indications that there
is no significant beaming in the low hard state.  We then consider in
\S 2.3 a particular source, 4U1543--47, in the high soft state and
again show that we can rule out any significant beaming.

\subsection{Low Hard State and Quiescent State}

Among the five common spectral states that black hole X-ray binaries
display (McClintock \& Remillard 2004), the low hard state and the
quiescent state are of particular interest.  These two states are
believed to be similar to each other in their underlying physics,
except that the quiescent state has a much lower luminosity and by
inference a lower mass accretion rate.  Both states are characterized
by the presence of a power-law spectral component in hard X-rays and a
near-absence of a blackbody-like component in soft X-rays (McClintock
\& Remillard 2004).  The latter component is expected to be dominant
if the source has a traditional optically thick disk; therefore, such
a disk is either absent or is energetically negligible.

In an important early paper, Shapiro, Lightman \& Eardley (1976)
described a hot optically-thin two-temperature accretion solution and
suggested that the low hard state of Cyg X--1 may correspond to this
solution.  However, Pringle (1976) showed that the solution is
thermally unstable.  Narayan \& Yi (1995) and Abramowicz et al. (1995)
demonstrated that there are, in fact, two distinct hot solutions: (i)
the solution discovered by Shapiro et al. (1976), and (ii) a second
two-temperature solution that is characterized by energy advection.
The latter solution, which was originally discovered by Ichimaru
(1977), is thermally stable.  It is named an advection-dominated
accretion flow (ADAF) and has been used for a number of years to model
the low and quiescent states of black hole binaries (Narayan,
McClintock \& Yi 1996; Esin et al. 1997, 1998, 2001; McClintock et
al. 2003; see Narayan, Mahadevan \& Quataert 1998 for a review).
According to this model, a black hole binary in the low or quiescent
state has a standard optically thick disk only at relatively large
radii, while the gas at smaller radii is in the form of an ADAF.  The
X-ray emission is produced by the ADAF via Comptonization of soft
photons.  

There are two sources of soft photons in the above model: synchrotron
emission from the hot electrons in the ADAF, and thermal photons from
the thin accretion disk on the outside (see Narayan, Barret \&
McClintock 1997; Esin et al. 1997).  Depending on the location of the
transition radius between the ADAF and the outer disk, one or the
other dominates the Comptonization.  Generally, synchrotron emission
dominates in the quiescent state and thermal radiation from the outer
disk dominates in the low hard state (and even more so in the
intermediate state between the low hard state and the high soft
state).

While the ADAF model provides a satisfactory description of the X-ray
emission in black hole X-ray binaries, it is unable to explain the
radio emission that is often observed in the low hard state.  This is
not surprising since the radio emission has been resolved into a jet
in Cyg X--1 and jets are inferred for other sources via a brightness
temperature argument (Fender 2004).  Markoff, Falcke \& Fender (2001)
proposed a model for XTE J1118+480 in which nearly the entire spectrum
from radio to hard X-rays is produced by synchrotron emission from a
jet.  Corbel et al. (2003) noted that there is a strong correlation
between the radio and X-ray emission in the black hole binary GX
339--4 in the low hard state and the quiescent state and suggested
that a significant fraction of the observed X-ray radiation in these
two states is produced by synchrotron emission from a jet.  In this
picture, the hot ADAF is virtually silent in all bands.

Several studies have followed up on this suggestion.  Heinz \& Sunyaev
(2003) worked out a scaling relation between the synchrotron flux at a
given frequency, the mass of the black hole, and the mass accretion
rate.  Their model is applicable to jets anchored in either an ADAF or
a standard disk.  Merloni, Heinz \& di Matteo (2003) extended this
work and showed that accreting black holes follow quite well a
``fundamental plane'' in the three-dimensional parameter space of
radio luminosity, X-ray luminosity, and black hole mass.  However, they
came down in favor of the ADAF rather than the jet as the source of
the X-ray emission in the low hard state.  Falcke, K\"ording \&
Markoff (2004) argued instead that synchrotron emission from the jet
is the source of the X-rays.  In a recent paper, Heinz (2004) has
presented additional arguments why a synchrotron jet is unlikely to
explain the X-ray emission in low hard state binaries.

In the context of spectral modeling, a jet has two important features.
First, the radiating particles are expected to be highly nonthermal;
this is usually modeled with a power-law distribution in energy.  A
nonthermal distribution is somewhat problematic for explaining the
X-ray emission since the hard X-ray spectra of several low hard state
binaries (e.g., Esin et al. 1998) show a turnover at about 100 keV.
This turnover is easily explained in the ADAF model since the thermal
electrons naturally acquire temperatures of this order (Esin et
al. 1997, 1998; Zdziarsky et al. 2003).  For a wide range of accretion
rates, the ADAF model gives electron temperatures that vary by only a
factor of a few.  The reason is that thermal Comptonization becomes
significantly more effective when the electrons are relativistic.
There is hence a natural thermostat in these models whereby the
electrons tend to equilibrate at $kT_e$ a factor of a few less than
$m_ec^2$.  Correspondingly, the Compton emission cuts off at
$\sim100-200$ keV.  In the jet synchrotron model, there is nothing
special about a photon energy of 100 keV, and so the observed turnover
of the spectra at this energy requires some degree of fine-tuning.

The second feature of a jet --- one that is unique to it --- is that
it involves relativistic bulk motion.  The jets seen in blazars, radio
galaxies and quasars often have bulk Lorentz factors $\gamma \sim 10$,
resulting in strong beaming of the radiation.  The radio jets seen in
some X-ray binaries in the very high state, the microquasars, have
$\gamma \sim 2.6$ (Mirabel \& Rodr\'iguez 1999), which again implies
considerable beaming.  If the X-ray emission in black hole binaries in
the quiescent and low hard states is from a jet, one might naively
expect the radiation to be strongly beamed.  However, recent X-ray and
radio studies of BHXN argue against such strong beaming in the low
hard state (Gallo, Fender \& Pooley 2003; Maccarone 2003).  In \S 2.2
we consider the quiescent state of these sources and investigate what
limits we can place on beaming.

\subsection{Black Hole X-ray Novae in Quiescence}

Table 1 lists the parameters of 10 well-studied BHXN in order of
increasing orbital period.  The inclination angles $i$ listed in
column 3 are taken for each source from the first reference cited.
The following reference or references support the distance and
luminosity data given in columns 4 and 5. The distances, apart from
GRO J1655--40 (Hjellming \& Rupen 1995), are based on a dynamical
model for each binary system that allows a determination of the
physical radius of the secondary star, and hence its absolute visual
magnitude, given its spectral type (e.g., Barret, McClintock \&
Grindlay 1996).  Taking into account a correction for interstellar
reddening and a correction for the continuum flux that is contributed
by the accretion disk, one obtains the apparent dereddened magnitude
of the secondary, and hence the distance.  The uncertainty in the
distances so obtained is typically $\pm25$\% (Barret et al. 1996).  We
note that these distances have been criticized as underestimates by
Jonker \& Nelemans (2004) on what we consider speculative grounds;
moreover, since those authors do not recommend a set of corrected
distances we have no choice but to use the standard distances listed
in Table 1.  The luminosity column gives for each source the lowest
(absorption-corrected) X-ray luminosity that has been reported in
observations made using either {\it Chandra} or {\it XMM-Newton}.

The values of inclination are of central importance to this work.  In
our selection of the data, we have favored values determined by
modeling IR light curves because they are less contaminated than
optical light curves by emission from the accretion disk.  However,
even the inclinations derived from IR light curves are subject to
criticism, as shown for example by the debate over the proper
interpretation of the light curves of A0620--00 (Shahbaz,
Bandyopadhyay \& Charles 1999; Froning \& Robinson 2001; Gelino,
Harrison \& Orosz 2001a).  Nevertheless, the spectral energy
distributions (BVRIJHK) obtained for some of the sources provide a
limit on the contamination and a convincing lower limit on the
inclination angle (e.g., Gelino et al. 2001a; Gelino \& Harrison
2003).  Moreover, in a few favorable cases such as GRO~J1655--40, the
nonstellar flux is a small fraction of the stellar flux (Greene,
Bailyn \& Orosz 2001).  Finally, the agreement between the
inclinations given in Table~1 and most of the earlier and less precise
determinations gives one reasonable confidence in the values we have
adopted (e.g., in the case of GS/GRS 1124--68, see Orosz et al. 1996,
and Shahbaz, Naylor and Charles 1997).

The upper panel in Figure 1 shows the quiescent X-ray luminosities of
the BHXN from Table 1 plotted against the inclinations $i$ of their
binary orbits.  The luminosities have been normalized by $10^{31}
~{\rm erg\,s^{-1}}$ for easy comparison with the normalized
theoretical curves.  The source GS 2023+338 (or V404 Cyg) has not been
shown in the plot.  This source has a subgiant companion and is much
brighter than the other BHXN; in fact, it is off-scale on the plot.
Fortunately, the source has a value of $\cos i\sim0.5$ (the
normalization point or pivot point of the theoretical curves), so it
does not bias the results.

We see from Figure 1 that there is no hint in the data for any
increase in the observed luminosity of low-inclination systems ($\cos
i \to 1$).  Indeed, the most pole-on system in the sample, 4U1543--47
with $i = 20.7\pm1.0\degr$, has a 95\% confidence upper limit on its
quiescent luminosity that is below the luminosity predicted for all
values of $\gamma$ (other than unity) considered in the plot.  The
lower panel in Figure 1 shows the same data binned in $\cos i$.  In
each bin, all the systems with measured luminosities in that bin have
been combined and their mean luminosity has been plotted, along with
the estimated error in the mean.  The systems with upper limits are
shown separately.  This plot confirms that there is no evidence for
any increase in the observed luminosity with increasing $\cos i$.  If
quiescent BHXN produce their X-ray emission from a jet, then it would
appear that the Lorentz factor of the jet must be fairly small.

Note the special importance of 4U1543--47.  With the lowest
inclination among all our sources, it has the greatest ability to test
for the presence of a jet.  The source has a relatively low
luminosity, but is this significant?  The answer is not obvious since
Figure 1 indicates that the quiescent luminosities of BHXN in general
have a large dispersion.  We attempt to provide a quantitative
analysis of this question here.

Leaving aside GS 2023+338 (see above), the seven objects in Figure 1
with measured quiescent luminosities have a mean X-ray luminosity of
$\log L_{\rm q}=31$ and a dispersion around the mean of
$\sigma_{\log{L}}=0.6$.  A fraction of the dispersion may be caused by
beaming, but most of the dispersion seems to be intrinsic to the
sources (see also Heinz \& Merloni 2004 for a related discussion).
For simplicity, and also to be maximally conservative, we assume that
the entire dispersion is intrinsic to the sources.  Now, given a value
for the jet Lorentz factor $\gamma$, and using the estimated
inclination angle of $i=20.7^o$ in 4U1543--47, we can estimate the
mean expected luminosity of the source $\langle\log L_{\rm
q}(\gamma)\rangle$.  (The theoretical curves in Fig. 1 show this
quantity for selected values of $\gamma$.)  Assuming a log-normal
probability distribution for $\log L_q$ with dispersion
$\sigma_{\log{L}}=0.6$, we then obtain the following probability
distribution for $\log L_{\rm q}$ of 4U1543--47:
\begin{equation}
P(\log L_{\rm q};~\gamma) = {1\over\sigma_{\log{L}}\sqrt{2\pi}}
\exp\left[ - {(\log L_{\rm q}-\langle\log L_{\rm
q}(\gamma)\rangle)^2\over 2\sigma_{\log{L}}^2}\right] d\log L_{\rm q}.
\end{equation}
This distribution is a function of $\gamma$ because the quantity
$\langle\log L_{\rm q}\rangle$ depends on it.

Garcia et al. (2001) observed 4U1543--47 for 9.9 ks with Chandra.
>From the estimated distance to the source, they state that a
luminosity of $L_{\rm q}=3.0\times10^{31} ~{\rm erg\,s^{-1}}$
corresponds to 5 counts, i.e., a luminosity of $L_{\rm
q}=6.0\times10^{30} ~{\rm erg\,s^{-1}}$ is equivalent to 1 count.  In
the actual observations, Garcia et al. detected no counts at all.
Assuming Poisson statistics, the probability of detecting zero counts
for a source luminosity $L_{\rm q}$ is equal to $\exp(-L_{\rm
q}/6.0\times10^{30})$.  Combining this with the probability
distribution for $\log L_{\rm q}$ written above, we obtain the
probability of measuring zero counts for a given jet $\gamma$ to be
\begin{equation}
P(0~{\rm cts};~\gamma) = {1\over\sigma_{\log{L}}\sqrt{2\pi}} \int
{\exp\left[ - {(\log L_{\rm q}-\langle\log L_{\rm
q}(\gamma)\rangle)^2\over 2\sigma_{\log{L}}^2} - {L_{\rm q}\over
6.0\times10^{30}} \right] d\log L_{\rm q}}.
\end{equation}
Doing the integral numerically, we find that the probability is 0.099
for $\gamma=1.24$, i.e., we can reject this value of $\gamma$ (or
anything larger) at the 90\% confidence level.  Similarly, we can
reject $\gamma\geq1.41$ at the 95\% confidence level.  These results
help to quantify the qualitative impression one obtains from Figure 1
that the quiescent luminosity of 4U1543--47 is inconsistent with
significant jet beaming.  Note that the upper limit on the quiescent
X-ray luminosity of 4U1543--47 was obtained with a relatively short
observation.  More sensitive X-ray observations could provide a much
more stringent limit.

Gallo, Fender \& Pooley (2003) obtained an upper limit of
$\gamma\sim2$ for the radio-emitting gas in the low hard state, while
Maccarone (2003) deduced $\gamma \lesssim 1.4$ for the X-ray-emitting
gas in GRO J1655--40.  To the extent that the quiescent state is
similar to the low hard state, the limit $\gamma\leq1.24$ in the present
study confirms these earlier indications.  Equally, however, the
evidence supports a jet-ADAF model for low-state and quiescent-state
BHXN.  In this model, the radio and other long-wavelength emission is
from a jet, but the high energy radiation, especially the hard X-rays,
is from an ADAF.  Malzac, Merloni \& Fabian (2004) and Yuan, Cui \&
Narayan (2004) have shown that such a model is consistent with the
spectral and timing data on XTE J1118+480 in the low hard state.

In jet models of the X-ray emission, it is proposed that both the
radio and X-ray emission are from the jet, though not from the same
region.  While the radio is emitted farther out by fast-moving gas,
the X-rays are postulated to come from the ``base of the jet'' where
the gas is moving more slowly.  It is not clear to the present authors
that the slowly-moving X-ray-emitting gas in these models is really
distinct from the hot gas in the underlying ADAF.  If the arguments
against a synchrotron origin of the X-ray emission from the jet
described in \S 2.1 are correct, then the jet model will need to
invoke Compton scattering to produce the X-rays.  The model would then
be so similar to the jet-ADAF model that it may be different only in
name.

While we have focused primarily on emission from a jet in the results
presented in Figure 1, it should be noted that even a pure accretion
model does involve moderately relativistic flow near the black hole.
However, the velocities are unlikely to exceed about $c/2$, and hence
$\gamma$ is at most $\sim1.1-1.2$.  The predicted effect due to
beaming is then too small to be distinguished with our present small
sample of sources.  Because of the different geometry, the beaming due
to an optically-thin axisymmetric accretion flow such as an ADAF will
give a larger luminosity as $\cos i \to 0$ compared to $\cos i \to 1$,
exactly opposite to the effect with a jet.  With a large enough sample
one might be able to identify this dependence.

Finally, we note that the inclination angles listed in Table 1 refer
to the binary orbit, which we have argued is likely to be coplanar
with the large-scale accretion disk.  However, none of the
measurements directly give the orientation of the jet.  In order to
obtain the latter, we made the additional assumption that the jet is
oriented perpendicular to the disk.  How reasonable is this?

Independent information on the jet orientation is available for two
sources, GRO J1655--40 and SAX J1819.3-2525.  Hjellming \& Rupen
(1995) estimated the jet angle in GRO J1655--40 to be $85\degr$, while
the binary inclination angle is measured to be $70\degr$ (Table 1).
The jet is thus misaligned from the disk normal by $\sim15\degr$,
which is a relatively small angle and has little effect on our
conclusions.  However, SAX J1819.3--2525 is more problematic.  In this
source, Orosz et al. (2001) interpreted the radio observations of
Hjellming et al. (2000) to deduce a jet angle $<10\degr$, compared to
a binary inclination of $75\degr$.  The misalignment between the jet
and the normal to the disk is thus more than $60\degr$, which is
enormous.  Maccarone (2002) has shown that the time scale on which the
spin axis of the black hole (which is presumably the same as the
direction of the jet) and the angular momentum axis of the disk become
aligned can be comparable to or longer than the lifetime of the binary
for transient sources with a low outburst duty cycle.  Therefore, such
a large misalignment could survive if it was somehow created initially
when the black hole was formed.

If the jet orientation in SAX J1819.3--2525 is indeed offset by a
large angle from the disk normal, then the conclusions reached in this
section regarding the presence/absence of jets and relativistic
beaming in the quiescent state of BHXN become less compelling.  (The
conclusions reached in \S 3 or \S 4 are not affected.)  We note,
however, that there are alternate explanations of the observations in
SAX J1819.3--2525 that do not require a small jet angle.  The reason
is that the radio observations of this source did not directly measure
the proper motion of the jet.  All that was measured was the angular
offset between the peak of the jet radio emission and the position of
the binary.  If the jet was ejected during the X-ray outburst that
occurred a short time prior to the radio observations, then the
position offset implies the jet angle quoted above: $<10\degr$.
However, as Chaty et al. (2001) have argued, the jet might have been
ejected a couple of weeks earlier than the X-ray outburst, in which
case the jet would have had a modest Lorentz factor and its
orientation would be consistent with the measured binary inclination.
This possibility becomes all the more reasonable considering that the
source showed optical activity for more than a month prior to the
X-ray outburst.

\subsection{4U1543--47 in the High Soft State}

Park et al. (2004) pointed out that, during the 2002 accretion
outburst of of 4U1543--47, the Fe line from the disk in this source
had an average equivalent width of $\sim300$ eV, which is three times
larger than the predicted equivalent width (George \& Fabian 1991) for
an unbeamed source.  Since it is unlikely that the theoretical
prediction is off by more than a factor of a few, Park et al. argued
that the amount of beaming allowed in the source is strongly
constrained.  Figure 2 quantifies this argument.

We assume that the hard X-rays irradiating the disk are produced in a
jet.  We then calculate the ratio of the X-ray flux density $S_{\rm
obs}$ seen by the observer at $i=21^{\rm o}$ (the orbital inclination
of 4U1543--47) to the mean flux density $S_{\rm irr}$ over the range
of angles $i=135^{\rm o}-180^{\rm o}$; the latter quantity is a
measure of the flux irradiating the disk, assuming that the
irradiation occurs over a cone of half-angle $45^{\rm o}$ around the
backward direction.  We take the energy spectral index to be $\alpha
=1.5$ in accordance with the observations of Park et al. (2004).  We
make the further reasonable assumption that the George \& Fabian
(1991) estimate of the Fe line equivalent width is correct to within
an order of magnitude, i.e., the equivalent width in the absence of
beaming is no larger than 1000 eV.  Relativistic beaming will cause
the equivalent width to decrease for an observer at low inclination.
This is because the observed hard X-ray continuum will increase as a
result of relativistic beaming, whereas the flux irradiating the disk
(and producing the fluorescent line) will decrease.  Since the
observed equivalent width is about 300 eV, the maximum value allowed
for the quantity $S_{\rm obs}/S_{\rm irr}$ in Figure 2 is (1000
eV)/(300 eV) $=10/3$.  This level is shown by the dotted line.

We see from Figure 2 that any X-ray emitting jet in 4U1543--47 in the
high soft state must have $\gamma<1.04$.  This is a rather tight
limit.  However, the result refers to a single source and should
perhaps not be taken as a general indication for all sources in the
high soft state.  Also, note that we are discussing here the high soft
state, which is very different from the low and quiescent states
discussed earlier.  The model of Cyg X--1 proposed by Beloborodov
(1999) and Malzac, Beloborodov \& Poutanen (2001), in which the hard
X-rays are produced in an outflowing corona, has a value of $\gamma$
roughly consistent with the limit derived here.

\section{On the Absence of Edge-On Systems}

An interesting feature of the sample of sources in Table 1 and Figure
1 is the absence of systems with $\cos i<0.25$, or $i>75^{\rm o}$.
Since $\cos i$ is proportional to the solid angle, it appears that
fully 25\% of the solid angle is for some reason not populated with
sources.  We quantify this argument here.

In addition to the 10 sources in Table 1, let us include also XTE
J1550--564, which does not have a quiescent luminosity measurement and
therefore is not listed in Table 1 but does have an estimated
inclination angle: $i = 72 \pm 5^{\rm o}$ (Orosz et al. 2002).  What
is the probability that these 11 sources would by chance all have
$\cos i > 0.25$?  The answer is $0.75^{11} = 0.042$.  That is, the
absence of sources with high inclinations ($\cos i < 0.25$,
$i>75\degr$) is significant at the 96\% confidence level, a fairly
strong result.  The conclusion does, however, require that our
estimates of inclination angles be reliable.

We now present an independent argument for the scarcity of highly
inclined systems.  This second argument does not require inclination
estimates but is based just on the fact that no black hole X-ray
binary has so far been found to eclipse.  If the X-ray-emitting inner
accretion disks in black hole binaries are visible from all
directions, then occasionally we ought to have a nearly edge-on binary
for which we should see X-ray eclipses.  The solid angle over which
eclipses will be visible is determined by the eclipse angle $\theta_e$
corresponding to a ray from the X-ray source that just grazes the top
(or bottom) of the secondary.  For Roche-lobe filling secondaries,
$\theta_e$ depends only on the mass ratio $Q\equiv M_{\rm BH}/M_s$,
where $M_{\rm BH}$ is the mass of the black hole and $M_s$ is the mass
of the secondary.  Table 2 lists all 20 dynamical black hole binaries
known today, along with estimates of the mass ratio $Q$.  The sources
are collected into four groups.

The first group in Table 2 consists of the 10 BHXN listed in Table 1
plus the BHXN GS1354--64.  These sources all have estimates of $Q$.
Using the given values of $Q$ we have used the Eclipsing Light Curve
code (version 2) described in Orosz \& Hauschildt (2000), which
incorporates full Roche geometry, to obtain the ranges of $\theta_e$
given in column 4.  Column 5 gives the mean eclipse angle
$\langle\theta_e\rangle$, which is the value of $\theta_e$
corresponding to the mean $Q$ from column 3.  (This is slightly more
conservative than using the mean $\theta_e$ from column 4.)  For each
binary, the probability that it will not be seen as an eclipsing
source is equal to $1-\cos\langle\theta_e\rangle$.  The mean value of
this quantity for the 11 sources is 0.792.  The probability that none
of the 11 would be an eclipsing source is the product of the
$1-\cos\langle\theta_e\rangle$ factors of the individual sources,
which gives 0.0766.

The second group in Table 2 lists two BHXN for which we have only
lower limits on $Q$.  If we conservatively take $Q$ to be 30 for these
two sources (a larger value than for any of the sources in the first
group), we obtain the estimated value
$\langle\theta_e\rangle=82.2\degr$ shown in parentheses in column 5.
The third group in Table 2 consists of four BHXN for which we have no
information on $Q$.  For these sources, we assume that their $Q$
values are similar to those of the first group and that their mean
value of $1-\cos\langle\theta_e\rangle$ is therefore 0.792.  This
leads to the estimated value $\langle\theta_e\rangle=77.8\degr$ given
in column 5.

Finally, the fourth group consists of three persistent sources (these
are not BHXN).  In view of the high X-ray luminosity of LMC X--3 and
the fact that an accretion disk is likely to be present in this
source, Kuiper, van Paradijs \& van der Klis (1988) argued that the
secondary probably fills its Roche lobe.  LMC X--1 is almost as
luminous, so this source too probably fills its Roche lobe.  We thus
obtain the values of $\langle\theta_e\rangle$ given in Table 2 for
these two sources.  Cyg X--1 probably underfills its Roche lobe, but
the degree to which it does so is uncertain.  Bolton (1975) considered
a radius range for the secondary from $0.9-1R_L$, where $R_L$ is the
radius of the Roche lobe, and Bolton (1986) considered the range
$0.8-1 R_L$.  To be conservative, we have assumed $0.8R_L$ in
calculating the value of $\langle\theta_e\rangle$.

Combining all 20 sources listed in Table 2, the probability that none
of them would be an eclipsing source is given by the product of the
individual $1-\cos\langle\theta_e\rangle$ values.  This gives 0.00535.
That is, the absence of eclipses is significant at the 99.5\% level.
This is a surprisingly strong result considering that we have assumed
nothing more than that the secondaries fill their Roche lobes (except
in the case of Cyg X--1, see above).  Note that the probability of
observing an eclipse varies like $Q^{-1/3}$ for large $Q$, so the $Q$
values given in Table 2 would need to be increased by a very large
factor before the absence of eclipses will be consistent with a normal
statistical fluctuation.

We thus have two quite independent arguments, both suggesting that the
inner accretion disks in black hole X-ray binaries are not equally
visible from all directions.  They must be completely hidden, or at
least strongly absorbed, when viewed nearly edge-on.  A likely
explanation is that the disks are flared by about $\pm 15^{\rm o}$ and
thus permanently occult the X-rays in more nearly edge-on systems.
The flare angle must be comparable to or larger than $\pi/2 -
\theta_e$.  This possibility was suggested many years ago by Milgrom
(1978) to explain the rarity of eclipses in neutron star X-ray
binaries.  We suggest that the same explanation applies also to black
hole systems.  Incidentally, we use the word ``flare'' here in a
generalized sense so as to include other geometries such as disk
``corrugation'' or a ``warp.''

\section{Dependence of Transient Light Curves on Inclination}

If edge-on systems are obscured from sight and face-on sources are
not, what about sources that are close to being obscured?  These would
be the sources with inclination angles in the range say
$70\degr-75\degr$ in Table 1.  Do these sources show any hint that
they are close to being obscured?  Here we focus on one signature: an
intriguing correlation between the inclination angle of BHXN and the
complexity of their outburst light curves.  In addition to the 10
sources in Table 1, we also include XTE J1550--564 with $i = 72^{\rm
o} \pm 5^{\rm o}$ (see \S3).

Figure 3 shows X-ray light curves of four sources: 4U1543--47,
$i=20.7\pm1.0\degr$; A0620--00, $i=40.8\pm3\degr$; GRO J1655--40,
$i=70.2\pm1.9\degr$, XTE J1550--564, $i=72\pm5\degr$.  Notice the
rather dramatic difference between the two low-inclination systems and
the two high-inclination systems.  The former sources have ``classic''
exponential light curves, whereas the latter sources have erratic and
more variable light curves.  Notice in particular the deep minimum
half-way through the light curves of GRO J1655--40 and XTE J1550-564,
which have no analog in the low-inclination systems.

The systems shown in Figure 3 are just representative examples, and the
reader is invited to look up the X-ray light curves of the remaining 7
systems listed in Table 1.  Tanaka \& Lewin (1995) and Chen, Shrader \&
Livio (1997) present pre-1996 light curves, and McClintock and Remillard
(2004) present {\it RXTE}-era light curves .  An inspection of these
light curves shows that the six systems with low or moderate inclination
(i.e., $i < 65^{\rm o}$: 4U1543--47, A0620--00, GRO~J0422+32,
GS/GRS~1124--683, GS~2023+338 and GS2000+25) display simple, classic
light curves that rise rapidly and decay exponentially.  Admittedly, GS
2023+338 exhibited extreme variability and peculiar behavior;
nevertheless, the upper envelope of its light curve resembles a classic
light curve (see Fig. 3.2 in Tanaka \& Lewin 1995).

We note that the low-inclination system 4U1543-47 has had three prior
outbursts, and there is no evidence for erratic behavior in any of the
earlier light curves.  Following the 1971 ouburst, the smooth
exponential decay phase was observed quite continuously for a period of
$\approx 120$ days (Li, Sprott \& Clark 1971).  The fast, smooth rise of
the 1983 outburst and the early decay period were closely observed for
about 8 days with the ASM aboard {\it Ginga} (Kitamoto et al. 1984).  No
relevant observations of the 1992 outburst have been reported (see Chen
et al. 1997).  Thus observations of two earlier outbursts of 4U1543-47
confirm the classic outburst profile shown in Figure 3.

In contrast to the above six sources, the other five systems that are
viewed at higher inclination angles near $70^{\rm o}$, namely
GRS~1009--45, XTE~J1118+480, GRO J1655--40, XTE J1550--564 and SAX
J1819.3--2525, all have more complex light curves that display
pronounced variability, large dips, and other unusual behavior.
Especially telling is the unprecedented light curve of the highest
inclination system, SAX~J1819.3--2525, with $i = 75 \pm 2^{\rm o}$. This
source was active at a very low level of intensity ($\sim 0.05$ Crab)
for almost a year before it rose abruptly to an intensity of 12.2 Crab.
Within two hours thereafter, the source returned to its pre-outburst
intensity (Wijnands \& van der Klis 2000).

The optical light curves of BHXN also follow a similar progression
from simple to complex as the inclination rises above about $70^{\rm
o}$.  We illustrate this fact by briefly comparing the well-sampled
optical/X-ray light curves of A0620-00 and XTE J1550-564.  A very
simple and satisfactory model of the UBV optical light curves of
A0620-00 ($i = 40.8\pm3\degr$), which is based on X-ray reprocessing
in the accretion disk, is given by Esin et al. (2000).  In contrast,
the optical light curves of XTE J1550--564 are inscrutable (Jain et
al. 2001).  In this high-inclination system ($i = 72\pm5\degr$) the
authors find no correspondence between the X-ray thermal accretion
disk component and the optical flux. However, the X-ray power-law flux
is anticorrelated with optical dip features, which motivates the
authors to hazard the exotic suggestion that the optical flux may be
up-scattered into the X-ray power-law component by the hot corona.

What might cause the correlation of light curve complexity with
increasing inclination in BHXN?  Obscuration of some sort is a
possible answer.  If we accept Milgrom's (1978) explanation of a
flared disk to understand the absence of high-inclination sources in
the observed sample (\S3), then it is not a big leap to think that the
angular extent of the obscuration may vary with time during the
outburst.  For instance, suppose the flare angle of the disk is
$\sim15\degr$ during the early and late parts of the outburst but
increases to $\sim20\degr$ in the middle of the outburst.  Then, for
systems with $i$ in the range $\sim70-75\degr$ we would have a dip in
the middle of the light curve whereas there would be little effect for
lower-inclination systems.  In addition, if the flaring angle has
short term fluctuations, it could lead to a noisy light curve for
high-inclination systems, versus a more smooth light-curve for
low-inclination systems.

In addition, the radiating gas may be at different heights above the
disk at different times during the outburst.  For instance, any
emission from an extended jet is less likely to be obscured by a
flared disk compared to emission from the surface of a thin disk.
This looks like a particularly attractive explanation for the strange
light curve of SAX J1819.3--2525.  We suggest that this system, with
the highest inclination angle of all our sources, was strongly
obscured most of the time but became bright during a large jet
ejection episode when the X-ray emitting gas was at a relatively large
height above the disk.  During the rest of the outburst, we probably
observed only scattered radiation from the outermost parts of the disk
corona, which might explain the very low flux level.  Such scattering
by a disk corona is observed for several eclipsing LMXBs that contain
neutron star primaries (e.g., see White \& Holt 1982; Parmar et
al. 1986).

Of course, this discussion is based on a small number of sources --- 6
with low inclinations and 5 with high inclinations --- and the
correlation may be a statistical fluke.  Interestingly, one could use
the same data to argue that the complexity of the light curve
increases with increasing binary orbital period.  Although the
evidence is not as good as for the correlation with inclination angle
that we have discussed, it is strong enough to take note.  A larger
sample will help to sort things out.

\section{Summary}

Starting with the available information on the orbital inclinations of
black hole X-ray binaries, we have reached several conclusions in this
paper.

\noindent
1. We find that the quiescent X-ray luminosities of BHXN show no trend
with inclination $i$ (Table 1, Fig. 1).  Comparing the observations
with the expected trend if the X-ray emission is from a jet oriented
perpendicular to the binary orbital plane, and considering in
particular the low-inclination source 4U1543--47, we conclude that the
bulk Lorentz factor $\gamma$ of the jet is $\leq1.24$ at the 90\%
confidence level and $\leq1.41$ at the 95\% confidence level.  This
result for the quiescent state agrees with a similar conclusion
reached by Gallo et al. (2003) for the radio-emitting gas and
Maccarone (2003) for the X-ray-emitting gas in the low hard state.

\noindent
2. The source 4U1543--47 has only an upper limit on its quiescent
X-ray luminosity.  Because of its low inclination angle
($i=20.7\pm1.0\degr$) this source is most sensitive to relativistic
beaming from a jet.  We urge greater efforts to measure the quiescent
luminosity of the source.  A detection at a level well below the
present upper limit could be problematic for any jet model that
invokes a reasonable velocity for the radiating gas.

\noindent
3. 4U1543--47 showed a strong fluorescent iron line with an equivalent
width of 300 eV in outburst in the high soft state.  Quantifying an
argument by Park et al. (2004), we show that the X-ray continuum must
have originated in gas with an outward bulk Lorentz factor $<1.04$
(Fig. 2).  If this gas was in a jet, then the jet must have been
highly sub-relativistic.  Alternatively, the X-rays could have come
from a slowly expanding corona (Beloborodov 1999; Malzac et al. 2001)
or from the disk.

\noindent
4. Of the 11 sources with estimates of $i$ (Table 1 plus the source
XTE J1550--564), we show that there are none with $i>75\degr$ or $\cos
i <0.25$.  The probability that this would happen by chance is only
4\%, suggesting that there is a selection effect that prevents edge-on
systems from being seen.

\noindent
5. None of the 20 binaries in our complete sample is an eclipsing
source.  The probability that it could happen by chance is only 0.5\%
(calculated based on the data given in Table 2).  This completely
independent argument, which moreover does not require estimates of
$i$, confirms that high inclination systems are hard to detect.

\noindent
6. In our view, the most plausible explanation for the absence of
eclipsing sources is that the accretion disks in black hole X-ray
binaries are flared (or corrugated or warped) by about $\pm15\degr$.
The disks thus permanently occult the X-ray emission for observers
located within $\pm15\degr$ of the binary orbital plane.  Milgrom
(1978) suggested the same explanation for the rarity of eclipses in
neutron star X-ray binaries.

\noindent
7. We show that there is a surprising dichotomy between the X-ray
light curves of BHXN with $i \lesssim 65\degr$ and those with
$i\sim70-75\degr$ (Fig. 3).  The former sources have smooth
well-behaved outburst light curves with exponential decays, whereas
the latter have noisy light curves, often with a deep minimum around
the mid-point of the outburst.  There are similar indications also in
the optical light curves.

\noindent
8. We tentatively suggest that the systems with $i\sim70-75\degr$ have
noisy light curves because they are close to being occulted by the
disk.  Fluctuations in the flaring angle of the disk and/or the height
of the X-ray emitting gas above the disk could cause the observed
variations in the flux.

\noindent
9. The source SAX J1819.3--2525 has the largest inclination angle
among all our sources: $i=75\pm2\degr$.  It also has the most extreme
X-ray light curve.  In this BHXN, we suggest that the primary X-ray
source was occulted almost throughout its outburst, with only a small
amount of radiation being observed from the outer regions of a large
corona.  However, for a period of just two hours, the source became
very bright in X-rays.  We suggest that during this brief episode
there was a powerful jet ejection; X-rays from the jet at a relatively
large distance from the black hole were observed directly without
being obscured.

\acknowledgments

We thank Jerry Orosz for advice on using the Eclipsing Light Curve
code, we acknowledge the use of the processed {\it SAS-3} data from
Kenneth Plaks, Jonathan Woo, and George Clark, and we thank Ron
Remillard for help with the ASM light curves in Figure 3.  This work
was supported in part by NASA grants NAG5-10780 and NAG5-9930 and by
NSF grant AST 0307433.

\begin{deluxetable}{ccccccccc}\label{tab:1}

\tablewidth{0pt} 

\tablecaption{Inclinations and Quiescent Luminosities of BHXN}
\tabletypesize{\small}
\tablehead{
  \colhead{System} & \colhead{$P_{\rm orb}$ (hr)} &
  \colhead{$i$ (degrees)} & \colhead{$D$ (kpc)} &
  \colhead{$\log[L_{\rm q} ({\rm erg\,s^{-1}})]$\tablenotemark{a}} &
  \colhead{References} }

\startdata

XTE J1118+480  &  4.1 & $70_{-7}^{+3}$ & 1.8 & 30.5 & 1,2 \\
				              	                
GRO J0422+32  &  5.1 & $45\pm2$ & 2.6 & 30.9 & 3,4 \\
 				              	                
GRS 1009--45  &  6.8 & 67(?) & 5.0 & $<30.9$\tablenotemark{b} & 5,6 \\
				              	                
A0620--00  &  7.8 & $40.8\pm3$ & 1.0 & 30.5 & 7,4 \\
				              	                
GS2000+25  &  8.3 & $64\pm1.3$ & 2.7 & 30.4 & 5,4 \\
				              	                
GS/GRS 1124--683  &  10.4 & $54_{-1.5}^{+4}$ & 5.5 & 31.6 & 8,9 \\
				              	                
4U1543--47 & 26.8 & $20.7\pm1.0$ & 7.5\tablenotemark{c} &
$<31.3$\tablenotemark{b} & 10,4 \\
				              	                
GRO J1655--40  &  62.9 & $70.2\pm1.9$ & 3.2 & 31.3 & 11,4 \\

SAX J1819.3--2525  &  67.6 & $75\pm2$ & 9.6 & 31.9 & 5,12,13 \\

GS 2023+338  &  155.3 & $56\pm4$ & 3.5 & 33.1 & 14,4

\enddata
\tablenotetext{a}{Quiescent luminosity in the 0.5--10 keV band.}
\tablenotetext{b}{Upper limits are at the 95\% level of confidence.}
\tablenotetext{c}{Distance obtained from reference 10.}
\tablerefs{
(1) J. Orosz et al. 2004, in preparation; (2) McClintock et al. 2003;
(3) Gelino \& Harrison 2003; (4) Garcia et al. 2001; (5) Orosz 2002; (6)
Hameury et al. 2003; (7) Gelino et al. 2001a; (8) Gelino et
al. 2001b; (9) Sutaria et al. 2002; (10) J. Orosz et al. 2004, in
preparation; (11) Greene et al. 2001; (12) Tomsick et al. 2003; (13)
Orosz et al. 2001; (14) Shahbaz et al. 1994.}

\end{deluxetable}

\newpage
\begin{deluxetable}{ccccccccc}\label{tab:2}

\tablewidth{0pt} 

\tablecaption{Grazing Eclipse Angles of Black Hole X-ray Binaries}
\tabletypesize{\small}
\tablehead{
  \colhead{System} & \colhead{$P_{\rm orb}$ (hr)\tablenotemark{a,b}} & 
  \colhead{$Q=M_{\rm BH}/M_s$\tablenotemark{a}} &
  \colhead{$\theta_e$ (degrees)} & \colhead{$\langle\theta_e\rangle$} &
  \colhead{References} }

\startdata

XTE J1118+480  &  4.1 & $22.7-28.8$ & $81.5-82.1$ & 81.8 & 1 \\
				              	                
GRO J0422+32  &  5.1 & $3.2-13.2$ & $74.6-79.9$ & 78.3 &  1 \\
 				              	                
GRS 1009--45  &  6.8 & $6.3-8.0:$ & $77.4-78.3$ & 77.9 & 1 \\
				              	                
A0620--00  &  7.8 & $13.3-18.3$ & $80.0-80.9$ & 80.5 & 1 \\
				              	                
GS2000+25  &  8.3 & $18.9-28.9$ & $81.0-82.1$ & 81.6 & 1 \\
				              	                
GS/GRS 1124--683  &  10.4 & $4.8-8.8$ & $76.3-78.6$ & 77.7 & 1 \\
				              	                
4U1543--47 & 26.8 & $3.2-4.0$ & $74.6-75.6$ & 75.1 & 1 \\

GS1354--64 & 61.1: & $6.7-9.1$ & $77.6-78.7$ & 78.1\tablenotemark{c} & 2 
\\

GRO J1655--40  &  62.9 & $2.4-2.7$ & $73.3-73.9$ & 73.6 & 1 \\

SAX J1819.3--2525  &  67.6 & $2.22-2.39$ & $73.0-73.3$ & 73.1 & 1 \\

GS 2023+338  &  155.3 & $16.1-18.9$ & $80.5-81.0$ & 80.8 & 1 \\

  &  &  &  &  &  \\

1H1705--250 & 12.5 & $>18.9$ & $>81.0$ & (82.2) & 1 \\

XTE J1550--564 & 37.0 & $>12$ & $>79.6$ & (82.2) & 1 \\

  &  &  &  &  &  \\

XTE J1650--500 & 7.7 & & & (77.8) & 3 \\

XTE J1859+226 & 9.2: & & & (77.8) & 1 \\

4U1658--48 & 42.1: & & & (77.8) & 4 \\

GRS 1915+105 & 804.0 & & & (77.8) & 1 \\

  &  &  &  &  &  \\

LMC X--3 & 40.9 & $1.1-2.0$ & $69.5-72.5$ & 71.2 & 1 \\

LMC X--1 & 101.5 & $0.3-0.7:$ & $62.3-67.0$ & 65.2 &  1 \\

Cyg X--1 & 134.4 & $0.50-0.57$ & $66.9-67.6$\tablenotemark{d} &
67.2\tablenotemark{d} & 1 &
				              	                
\enddata
\tablenotetext{a}{Colon denotes uncertain value.}
\tablenotetext{b}{For orbital period data see McClintock \& Remillard
(2004) and references 2, 3 \& 4.}
\tablenotetext{c}{Based on the best-fit value for the mass ratio, 
$Q=7.7$.}
\tablenotetext{d}{The radial fraction of the Roche lobe that is filled
  by the star is assumed to be 0.8 (see text).}
\tablerefs{
(1) Orosz 2002; (2) Casares et al. 2004; (3) Orosz et al. 2004; (4)
  Hynes et al. 2003.}

\end{deluxetable}

\newpage
\begin{figure}
\figurenum{1}
\plotone{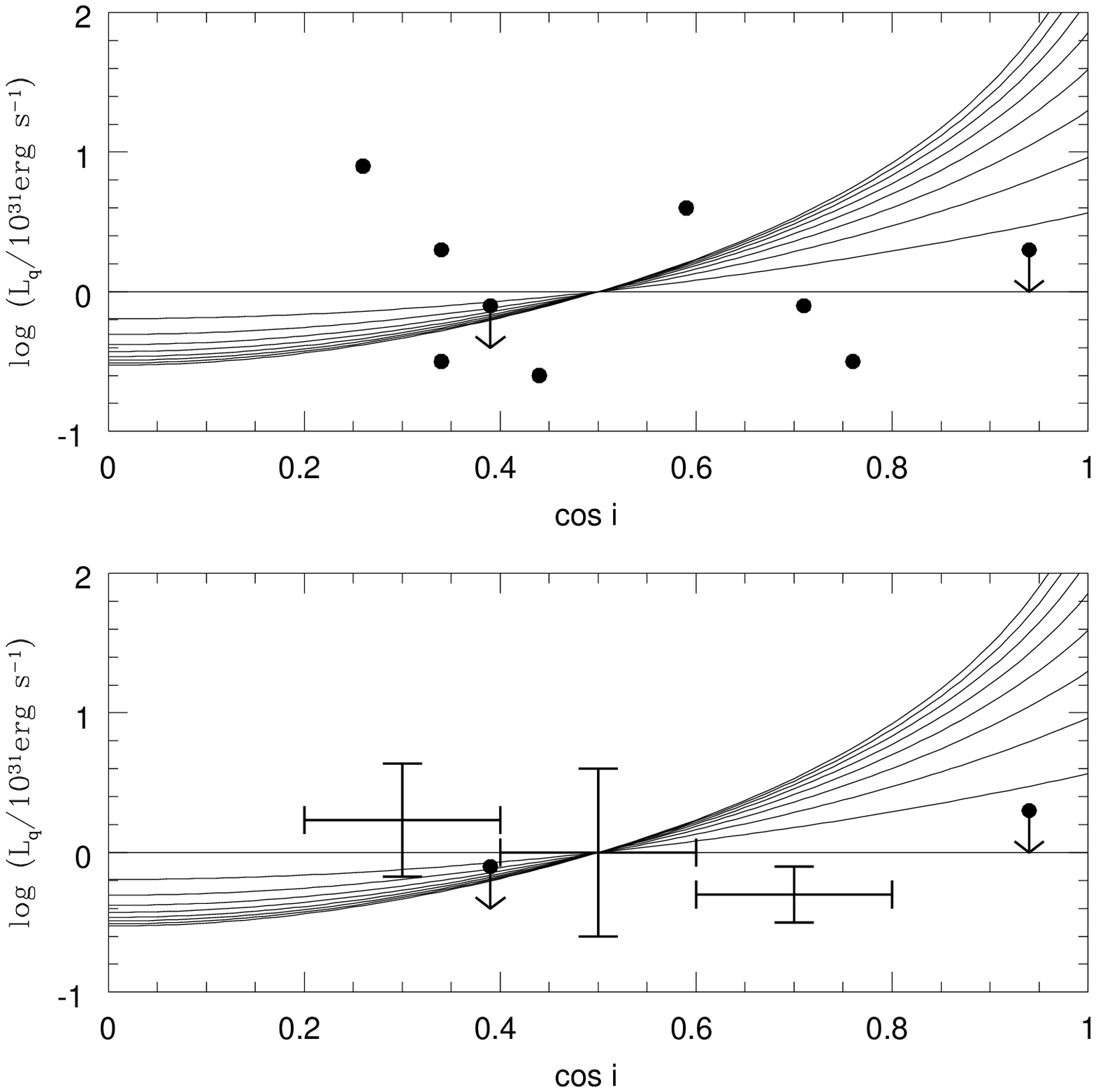}
\caption{Upper panel: Quiescent X-ray luminosities of BHXN in units of
$10^{31} ~{\rm erg\,s^{-1}}$ plotted against the cosine of the
inclination angle of the binary.  The curves indicate the expected
variation according to a jet model for different choices of the jet
Lorentz factor: $\gamma = 1.0$ (horizontal line), 1.2, 1.4, ... , 2.6.
Lower panel: The same data grouped into bins of width 0.2 in $\cos i$.
Upper limits are shown separately.  Both panels suggest that there is
no significant relativistic beaming of the X-ray emission in quiescent
BHXN. }
\end{figure}

\newpage
\begin{figure}
\figurenum{2}
\plotone{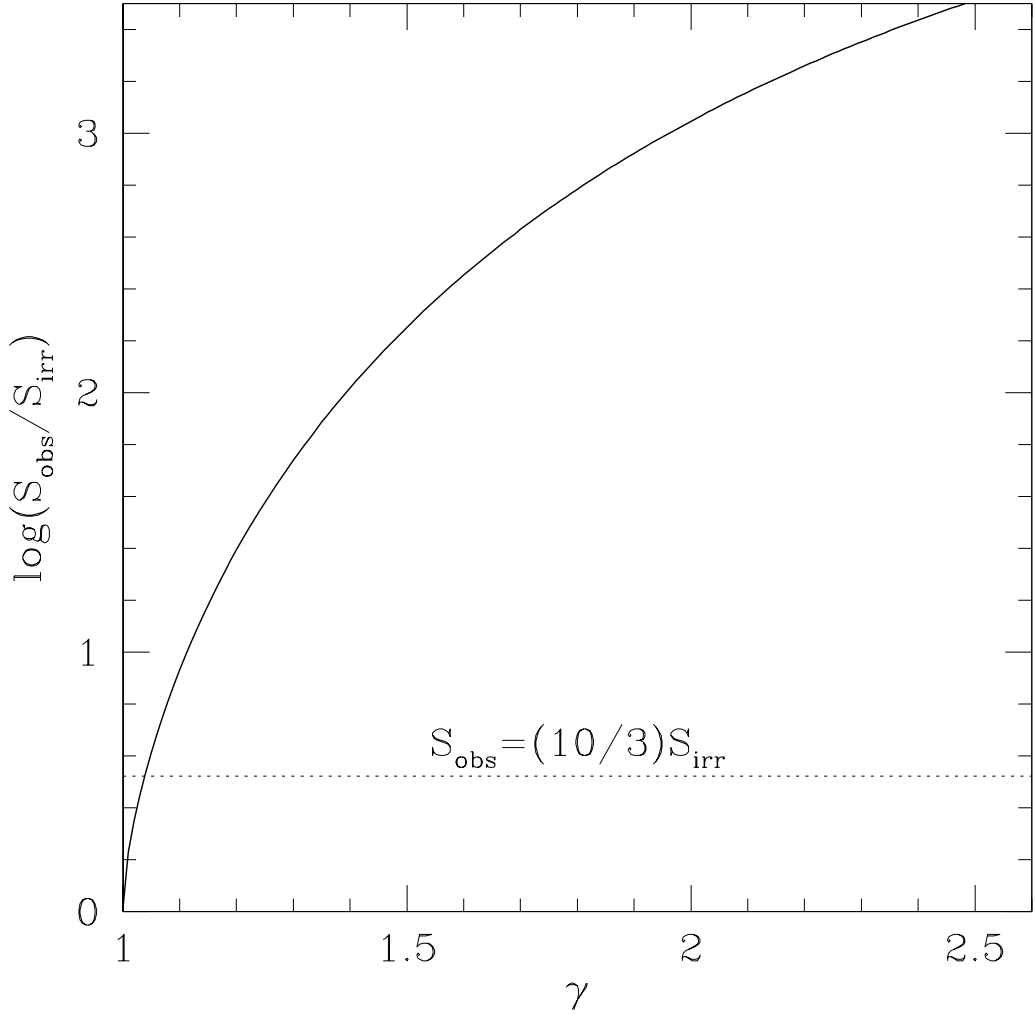}
\caption{Ratio of X-ray flux density from a jet in 4U1543--47 towards
the observer to the mean flux density towards the accretion disk,
plotted as a function of the jet Lorentz factor $\gamma$.  The
horizontal dotted line shows the upper limit on the ratio as estimated
from the equivalent width of the Fe line.  If the hard X-rays in
4U1543--47 during outburst are produced in a jet, then $\gamma$ must
be close to unity. }
\end{figure}

\newpage
\begin{figure}
\figurenum{3}
\plotone{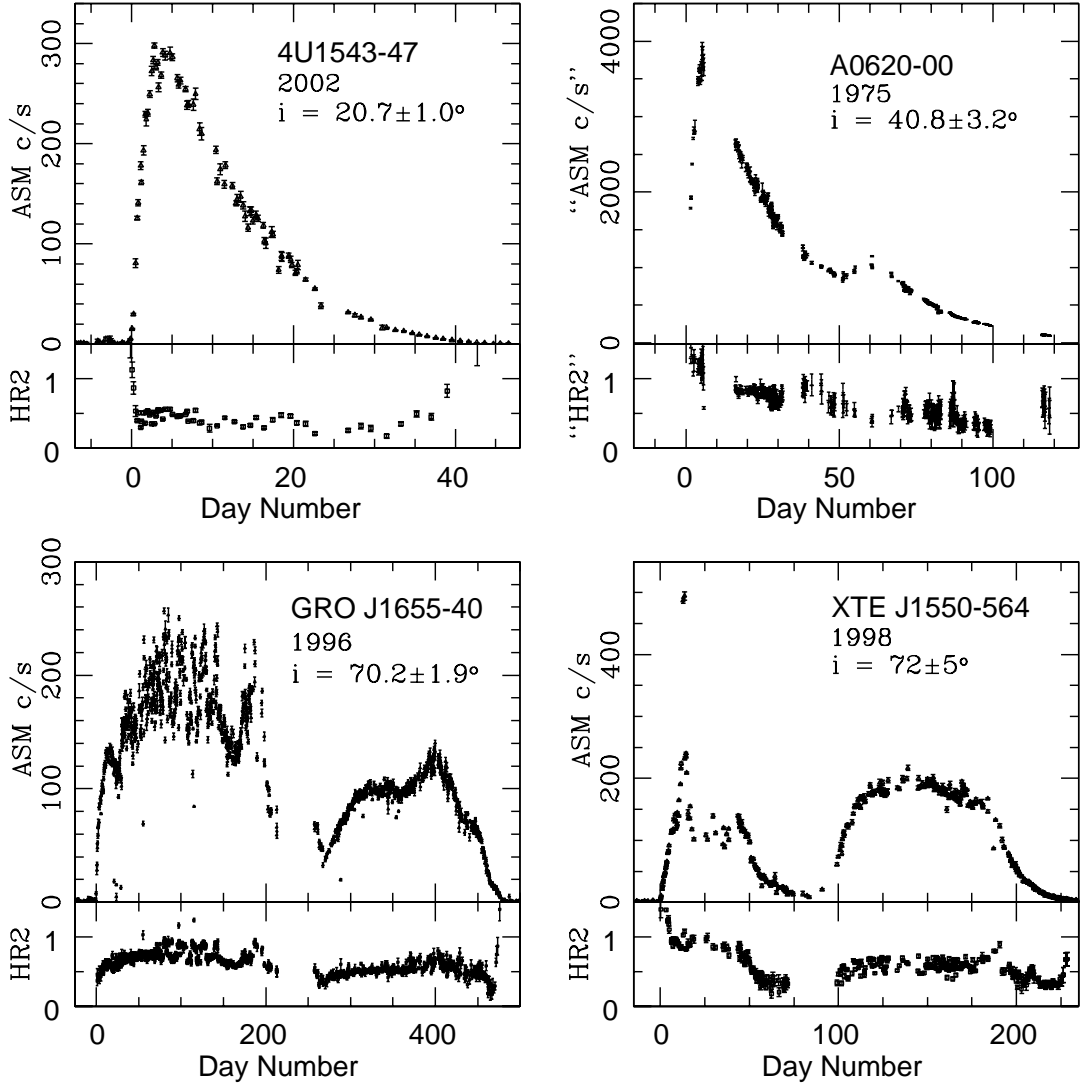}
\caption{Simple X-ray light curves observed for BHXN at low inclination
  (top two panels) are compared here to the complex light curves
  observed near $i \sim 70^{\rm o}$ (bottom two panels).  The more
  recent outbursts of three of the sources (4U1543--47, XTE J1550--564,
  and GRO J1655--40) were observed with the All-Sky (ASM) monitor aboard
  {\it RXTE}.  The light curves are for the 1.5--12 keV band and the
  hardness ratio HR2 is the 5--12 keV intensity divided by the 3--5 keV
  intensity.  An intensity of 1 Crab corresponds to 75.5 ASM c~s$^{-1}$.
  Time zero is the discovery date of the outburst in question.  The
  1.5--15 keV light curve of A0620--00 was obtained with the ``center
  slat'' detector aboard {\it SAS-3}, and the ratio of the 6--15 keV to
  the 1.5-6 keV intensities defines the {\it SAS-3} hardness ratio (Buff
  et al. 1977).  These quantities have been normalized to correspond
  approximately with the {\it RXTE} values of ASM intensity and hardness
  ratio.}
\end{figure}

\end{document}